\documentclass[%
 reprint,
 amsmath,amssymb,
 aps,
]{revtex4-2}

\usepackage{graphicx}
\usepackage{dcolumn}
\usepackage{bm}
\usepackage[utf8]{inputenc}

\usepackage{xcolor}
\usepackage{booktabs}

\begin{document}

\title{Flavor-Dependent QCD Critical Endpoint and Dual-Channel Fluctuations from Multi-Charge Holography}

\author{Zhibin Li$^{1}$}
\email{lizhibin@zzu.edu.cn}
\author{Danning Li$^{2}$}
\email{lidanning@jnu.edu.cn}
\author{Mei Huang$^{3}$}
\email{huangmei@ucas.ac.cn}

\affiliation{ $^{1}$ Institute for Astrophysics, School of Physics, Zhengzhou University, Zhengzhou 450001, China }
\affiliation{ $^{2}$ Department of Physics and Siyuan Laboratory, Jinan University, Guangzhou 510632, China}
\affiliation{ $^{3}$ School of Nuclear Science and Technology, University of Chinese Academy of Sciences, Beijing 100049, China   
         }


\begin{abstract}
We construct a thermodynamically self-consistent holographic QCD framework incorporating multiple conserved charges. By introducing three independent bulk $U(1)$ gauge fields, our Einstein-Maxwells-dilaton (EMsD) model naturally accommodates the coupled chemical potential landscape $(\mu_B, \mu_Q, \mu_S)$ inherent to realistic heavy-ion collisions. Crucially, thermodynamic consistency is enforced at the level of holographic renormalization, ensuring exact Maxwell cross-derivative relations without ad hoc patching. Calibrated exclusively at zero density, the model exhibits genuine predictive power for finite-density thermodynamics. We reveal that finite charge and strangeness densities induce pronounced nonmonotonic shifts in the critical endpoint (CEP) location. Furthermore, by mapping the freeze-out trajectories, we demonstrate that the allowed parameter bands robustly encompass empirical hadron resonance gas (HRG) fits. Within this physical regime, higher-order cumulant ratios for both net-baryon and net-charge channels exhibit coherent critical peaks at $\sqrt{s_{NN}} \approx 5\text{--}7\,\text{GeV}$. This hierarchical dual-channel signature provides a decisive, background-free strategy for the ongoing experimental search for the QCD critical point.
\end{abstract}

\maketitle

\textit{Introduction.}---
Mapping the QCD phase diagram at finite temperature ($T$) and baryon chemical potential ($\mu_B$) is central to understanding strong-interaction QCD matter, with direct implications for relativistic heavy-ion collisions and compact-star physics. While lattice QCD has precisely determined the crossover transition and equation of state at $\mu_B=0$~\cite{Borsanyi:2010cj,Borsanyi:2013bia,HotQCD:2014kol}, the fermion sign problem precludes direct simulations at finite density~\cite{deForcrand:2009zkb}, blocking first-principles access to the intermediate-density regime probed by the RHIC Beam Energy Scan (BES)~\cite{Garg:2013ata,Luo:2017faz,Andronic:2009jd, Andronic:2017pug}. Consequently, complementary nonperturbative approaches are indispensable: effective chiral models (e.g., NJL/PNJL) capture chiral symmetry breaking and deconfinement dynamics~\cite{Klevansky:1992qe,Fukushima:2004qy,Asakawa:1989bq,Schwarz:1999dj,Li:2018ygx,Zhuang:2000ub}, while functional methods like Dyson-Schwinger equations and the functional renormalization group provide sign-problem-free access to finite-density QCD~\cite{Xin:2014ela,Gao:2016qkh,Qin:2010nq,Shi:2014zpa,Fischer:2014ata,Gao:2020qsj,Fu:2019hdw,Zhang:2017icm,Fu:2021oaw,Fu:2022gou}. The holographic Einstein-Maxwell-Dilaton (EMD) framework has emerged as a powerful tool for studying QCD matter~\cite{DeWolfe:2010he,Cai:2012xh,Cai:2012eh,Finazzo:2013efa,Li:2017ple,Chen:2017cyc,Knaute:2017opk,Li:2020hau,Grefa:2021qvt,He:2022amv,Chen:2024ckb}. Driven by advanced numerical techniques, particularly machine learning, this approach now achieves remarkable quantitative consistency with lattice QCD, providing a robust theoretical foundation for locating the QCD critical end point (CEP)~\cite{Cai:2022omk, Hippert:2023bel, Chen:2024ckb, Luo:2024iwf, Zhu:2025gxo, Sun:2025uga}.

A primary objective is locating the QCD critical end point (CEP), which connects the low-$T$ first-order transition to the high-$T$ crossover. Near the CEP, critical fluctuations dominate thermodynamic observables only within a narrow parameter vicinity~\cite{Schaefer:2006ds}, yielding characteristic nonmonotonic structures in conserved-charge cumulant ratios ($C_n$)~\cite{Asakawa:2009aj,Schaefer:2011ex,Fan:2016ovc,Portillo:2016fso,Fan:2017mrk,Li:2017ple,Li:2018ygx,Fu:2021oaw,Zhao:2023xpj}. Experimentally, the STAR chemical freeze-out curve overlaps the QCD crossover for $\mu_B/T\le 3$~\cite{STAR:2017sal}. If the CEP lies at moderately larger $\mu_B/T$, realistic freeze-out trajectories can approach the critical region, imprinting these nonmonotonic fluctuation signatures on measurable net-proton cumulants~\cite{Fan:2016ovc,Fan:2017mrk,Li:2017ple,Portillo:2017gfk,Li:2018ygx,Wang:2018sur,Fu:2021oaw,Zhao:2023xpj,Huang:2023ogw,Stephanov:2011pb}.

However, to date, the existence of the CEP remains unconfirmed by both experimental and theoretical investigations. Lattice QCD excludes a CEP for $\mu_B/T \leq 3$ and $\mu_B < 300~\text{MeV}$~\cite{Vovchenko:2017gkg,Borsanyi:2020fev,Bazavov:2020bjn,Borsanyi:2021sxv,Bollweg:2022fqq,Philipsen:2021qji}, and recent STAR net-proton data show no clear nonmonotonic CEP signal for $\sqrt{s_{NN}} > 7.7$ GeV, with the key observable $\kappa\sigma^2$ exhibit only approximately monotonic behavior~\cite{STAR:2025zdq}. Critically, most existing theoretical calculations rely on idealized thermodynamic setups considering only $\mu_B$, neglecting the coupled effects of finite electric ($\mu_Q$) and strange ($\mu_S$) chemical potentials that are intrinsic to realistic heavy-ion collisions. This omission introduces substantial theoretical uncertainties, as flavor-dependent thermodynamics can significantly reshape both the CEP location and critical fluctuation patterns.

In this Letter, we address this gap using a five-dimensional Einstein-Maxwells-dilaton (EMsD) holographic model that naturally incorporates multiple flavor-resolved gauge fields. Rigorously calibrated to continuum-extrapolated lattice QCD data at vanishing density, the model provides robust, parameter-free predictions at finite density while strictly satisfying thermodynamic consistency across all three conserved charges ($B, Q, S$). By scanning the $(\mu_Q, \mu_S)$ parameter space under realistic freeze-out conditions (strangeness neutrality $\rho_S=0$ and charge-to-baryon ratio $\rho_Q/\rho_B=0.4$), we reveal a strong nonmonotonic flavor dependence of the CEP, where $\mu_B^{\text{CEP}}$ can shift continuously up to $\sim 600~\text{MeV}$. Crucially, our analysis demonstrates unified critical peak structures in multiple higher-order cumulant ratios ($\chi_3^B/\chi_1^B$, $\chi_4^B/\chi_2^B$, $\chi_4^Q/\chi_2^Q$) within a narrow collision-energy window of $\sqrt{s_{NN}} \approx 5\text{--}7~\text{GeV}$. These peaks exhibit a hierarchical amplitude pattern reflecting the intrinsic multi-charge dynamics of QCD criticality. These findings establish a dual-channel observational strategy for ongoing CEP searches and underscore that coupled multi-chemical-potential effects are indispensable for accurately mapping the QCD phase structure.

\textit{The EMsD Model and Thermodynamic Consistency.}---
To describe hot and dense QCD matter with multiple conserved charges, we adopt a five-dimensional Einstein-Maxwells-dilaton (EMsD) holographic model. The gravitational action reads
\begin{eqnarray}
    S=\frac{1}{2\kappa_5^2}\int d^5x \sqrt{-g}\left[\mathcal{R}-\frac{1}{2}\nabla_{\mu}\phi\nabla^{\mu}\phi-V(\phi) \right. \nonumber \\ 
    \left. -\sum_{q=u,d,s}\frac{Z_q(\phi)}{4}F_q^{\mu\nu}F_{q\mu\nu}\right],
\end{eqnarray}
where $\kappa_5^2=3.36\pi$ is the five-dimensional gravitational coupling, $g$ is the metric determinant, $\mathcal{R}$ is the Ricci scalar, and $\phi$ is the dilaton field encoding the QCD scale dependence. The term $F_q^{\mu\nu}$ denotes the field strength tensor for the multiple Maxwell gauge fields ($q=u,d,s$), while the flavor-dependent function $Z_q(\phi)$ introduces nonminimal dilaton-gauge couplings essential for reproducing flavor-differentiated thermodynamics.

To achieve quantitative agreement with lattice QCD at zero density, we parametrize the dilaton potential as \cite{Li:2025lmp}
\begin{equation}
V(\phi)=-12 \cosh\left(c_1 \phi\right)+\left(6c_1^2-\frac{3}{2}\right)\phi^2+c_2\phi^6 +c_3 \phi^8,
\end{equation}
with $c_1=0.71$, $c_2=3.7\times 10^{-3}$, and $c_3=2.8\times 10^{-5}$. The dilaton source is set to $\phi_s=1085$ MeV, and the renormalization coefficient is $b=-0.27435$ \cite{Li:2025lmp}. The gauge coupling functions take the form
\begin{equation}
Z_q(\phi)= \left(\cosh\left[ c_{q1}~\phi^{c_{q2} }\right]+c_{q3}\phi^2\right)^{-1},
\end{equation}
where the parameters $c_{qi}$ are calibrated separately for light ($u,d$) and strange ($s$) quarks, as summarized in Table~\ref{tab:params}.

\begin{table}[htbp]
\centering
\caption{Parameters of the $2+1$ flavor EMsD model calibrated to zero-density lattice QCD data. The values for the flavor-dependent couplings are listed separately for light ($u,d$) and strange ($s$) quarks.}
\label{tab:params}
\begin{tabular}{ccc}
\toprule
Parameter & Light quarks ($u,d$) & Strange quark ($s$) \\\midrule
$c_{q1}$ & 0.042 & 0.34 \\
$c_{q2}$ & 3.7 & 1.97 \\
$c_{q3}$ & -0.04 & -0.06 \\\bottomrule
\end{tabular}
\end{table}

In the holographic dictionary, the bulk dilaton $\phi$ is dual to the boundary running coupling (or equivalently, the trace anomaly operator), while the three independent bulk $U(1)$ gauge fields correspond to the conserved vector currents of $u$, $d$, and $s$ quarks. Equilibrium thermodynamic observables are derived via standard holographic renormalization. The higher-order susceptibilities of conserved charges are defined as 
\begin{eqnarray}
    \chi_n^M=\frac{\partial^n (P/T^4)}{\partial (\mu_M/T)^n}, \quad \chi_{mn}^{MN}=\frac{\partial^{m+n} (P/T^4)}{\partial (\mu_M/T)^m\partial (\mu_N/T)^n},
\end{eqnarray}
where $P$ is the pressure, $T$ is the temperature, and $\mu_M, \mu_N$ are the chemical potentials for charges $M, N \in \{B, Q, S\}$. By construction, these observables strictly satisfy the Maxwell cross-derivative relations ($\partial^2 P / \partial \mu_M \partial \mu_N = \partial^2 P / \partial \mu_N \partial \mu_M$). This guarantees that the holographic black hole thermodynamics first law is rigorously mapped to the boundary grand canonical ensemble, providing a solid foundation for analyzing critical fluctuations \cite{Li:2025lmp}.

\begin{figure*}[tbp]
\centering
\includegraphics[width=0.30\textwidth]{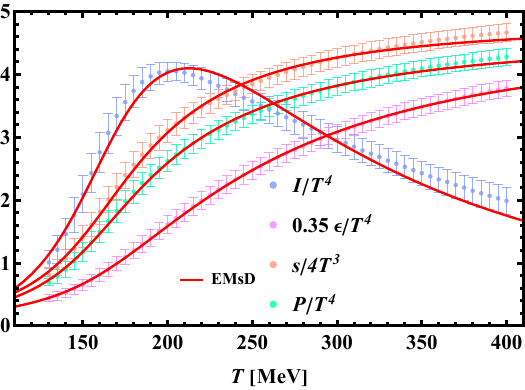}
\hfill
\includegraphics[width=0.315\textwidth]{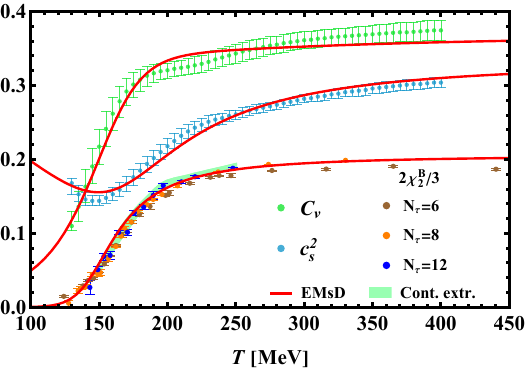}
\hfill
\includegraphics[width=0.315\textwidth]{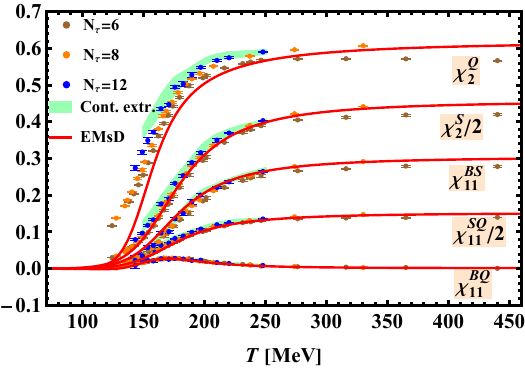}
\caption{Temperature dependence of equilibrium thermodynamic quantities at $\mu_Q=\mu_S=0$, comparing EMsD predictions with continuum-extrapolated lattice QCD results \cite{HotQCD:2012fhj, HotQCD:2014kol, Bazavov:2017dus}.}
\label{fig:1}
\end{figure*}

\textit{Equation of State and Finite-Density Predictive Power.}---
We systematically validate the EMsD model EoS against continuum-extrapolated lattice QCD data at vanishing chemical potentials ($\mu_B=\mu_Q=\mu_S=0$). A comprehensive comparison of key equilibrium observables across the QCD crossover region is presented in Fig.~\ref{fig:1}.

Fig.~\ref{fig:1} is organized as follows: \textit{Left panel:} fundamental bulk thermodynamics, including scaled pressure $P/T^4$, energy density $\epsilon/T^4$, entropy density $s/T^3$, and the trace anomaly $I/T^4$ (where $I = \epsilon - 3P$). \textit{Middle panel:} medium response functions, including the squared speed of sound $c_s^2$, volume heat capacity $C_V$, and second-order baryon susceptibility $\chi_2^B$. The pronounced minimum of $c_s^2$ near the crossover temperature $T_c$ signifies the softening of the EoS, which is of crucial importance for the hydrodynamic evolution and collective flow observables in heavy-ion collisions. \textit{Right panel:} second-order susceptibilities and mixed cross-correlations for electric and strange charges, quantifying the thermal fluctuation properties of the quark-gluon plasma. The EMsD predictions align with lattice QCD data within statistical uncertainties, verifying the framework at zero density.

\begin{figure*}[tbp]
\centering
\includegraphics[width=0.31\textwidth]{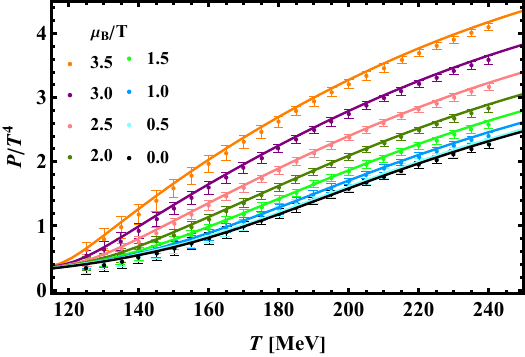}
\hfill
\includegraphics[width=0.31\textwidth]{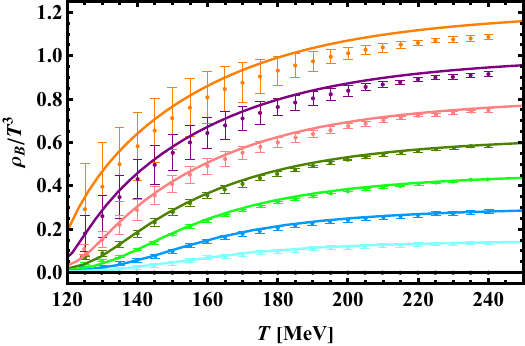}
\hfill
\includegraphics[width=0.31\textwidth]{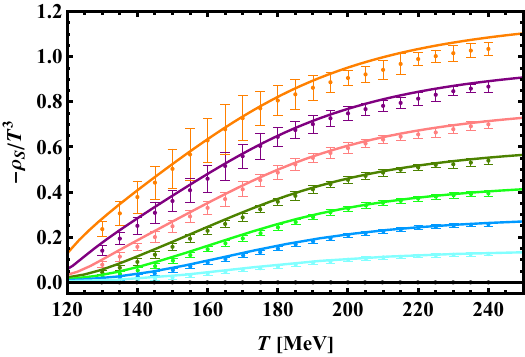}\\
\includegraphics[width=0.31\textwidth]{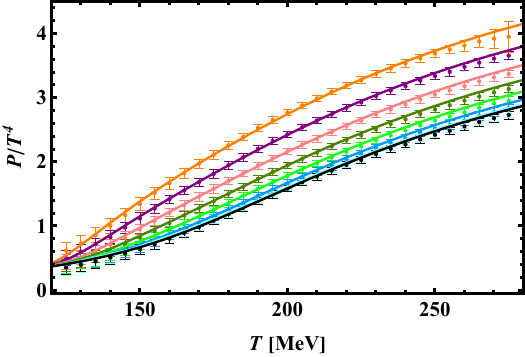}
\hfill
\includegraphics[width=0.31\textwidth]{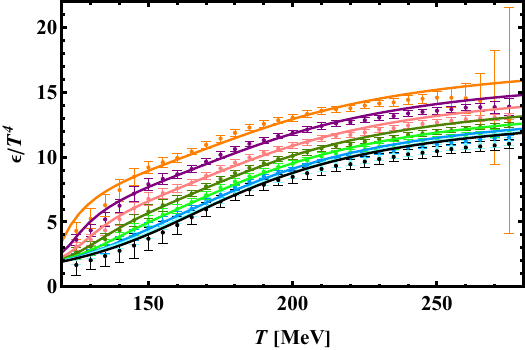}
\hfill
\includegraphics[width=0.31\textwidth]{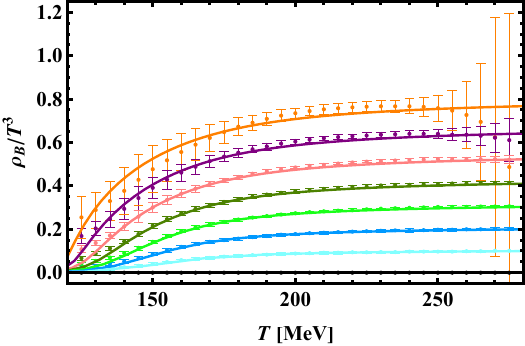}
\caption{Equilibrium thermodynamic observables along fixed $\mu_B/T$ trajectories in the finite-density regime. Upper row: $\mu_Q=\mu_S=0$. Lower row: $\mu_Q=0, \rho_S=0$ (strangeness neutrality). All EMsD results are compared with lattice QCD data \cite{Borsanyi:2021sxv,Borsanyi:2022qlh}.}
\label{fig:2}
\end{figure*}

To establish validity at finite baryon density, Fig.~\ref{fig:2} compares EMsD observables with lattice QCD Taylor expansion results along fixed $\mu_B/T$ trajectories. \textit{Upper row:} pressure, baryon density $\rho_B$, and strange density $\rho_S$ for charge- and strange-neutral conditions ($\mu_Q=\mu_S=0$). \textit{Lower row:} scaled pressure, energy density, and $\rho_B$ for strangeness-neutral conditions ($\mu_Q=0, \rho_S=0$). Physically, the $\mu_Q=\mu_S=0$ condition corresponds to idealized symmetric systems, while the $\rho_S=0$ condition strictly enforces local strangeness conservation, which is more relevant to the local fireball evolution in realistic heavy-ion collisions. Both span $0 \le \mu_B/T \le 3.5$. Notably, the EMsD predictions agree with lattice data within reported error bands under both conditions without introducing any additional finite-density parameters. This simultaneous consistency demonstrates that the flavor-decoupled parametrization $Z_q(\phi)$ is highly robust, and the model possesses genuine predictive power for the core thermodynamics of hot, dense QCD matter.

\begin{figure}[tbp]
\centering
\includegraphics[width=.23\textwidth]{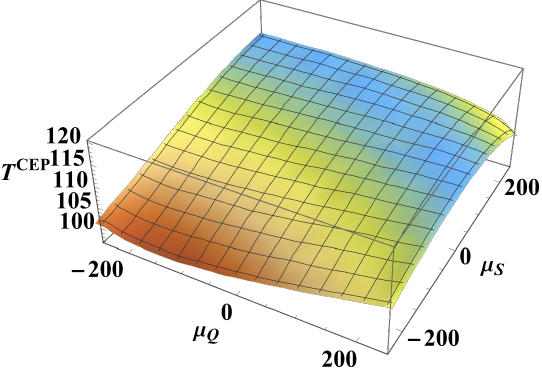}\hfill
\includegraphics[width=.23\textwidth]{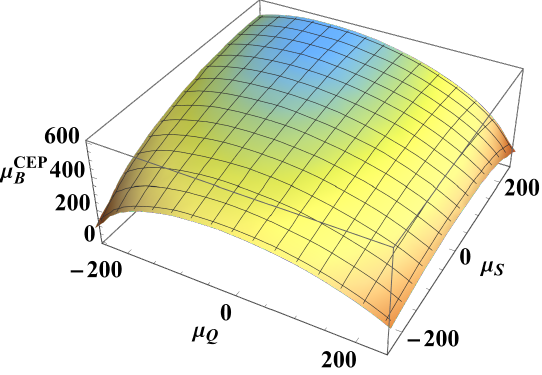}
\caption{CEP temperature $T_{\text{CEP}}$ (left) and CEP baryon chemical potential $\mu_B^{\text{CEP}}$ (right) over the scanned parameter range $-270\,\text{MeV}\le\mu_Q,\mu_S\le270\,\text{MeV}$, exhibiting nonmonotonic flavor dependence.}
\label{fig:3}
\end{figure}

\textit{Flavor-Dependent CEP Landscape and Critical Fluctuations at Freeze-out.}---
Having validated the model, we explore the flavor dependence of the QCD critical endpoint (CEP) by varying electric and strange quark chemical potentials. Scanning $-270\,\text{MeV}\le\mu_Q,\mu_S\le270\,\text{MeV}$, we map the continuous evolution of CEP coordinates in Fig.~\ref{fig:3}. The CEP temperature $T_{\text{CEP}}$ shows mild yet distinct nonmonotonic variations: lower values occur in the negative ($\mu_Q,\mu_S$) quadrant, while positive values elevate $T_{\text{CEP}}$. In contrast, $\mu_B^{\text{CEP}}$ displays strong nonmonotonic sensitivity, varying continuously from $0$ to $\sim 600\,\text{MeV}$. The maximum $\mu_B^{\text{CEP}}$ occurs at negative $\mu_Q$ and positive $\mu_S$. 

The pronounced shifts of the CEP location originate from the distinct roles of different quark flavors in the chiral phase transition. Due to its larger mass, the strange quark has a higher deconfinement and chiral restoration temperature scale compared to the light $u,d$ quarks. Variations in $\mu_S$ significantly alter the effective degrees of freedom and the thermodynamic feedback on the baryon density, thereby strongly displacing the CEP along the $\mu_B$ axis. These results highlight that multi-flavor chemical potentials are indispensable for precisely determining the QCD phase diagram structure.

To connect theoretical predictions to heavy-ion measurements, we constrain the EMsD model to realistic chemical freeze-out conditions. We adopt empirical parametrizations for the collision-energy-dependent freeze-out temperature and baryon chemical potential:
\begin{eqnarray}
\begin{split}
     \mu_B&=\frac{a}{1+b \sqrt{s_{NN}}}+\mu_{\Delta}, \\
     T&=\frac{T_{\text{lim}}}{1+\exp\left[ c-\ln{\left(d \sqrt{s_{NN}}+e\right)/0.45 } \right]},
\end{split}
\end{eqnarray}
where $\sqrt{s_{NN}}$ is the center-of-mass energy per nucleon pair. The parameters $a, b, c, d, e$ are empirical fit coefficients, $T_{\text{lim}}$ is the limiting temperature at high energies, and $\mu_{\Delta}$ accounts for fitting uncertainties. The parameter ranges, constrained by optimal hadron resonance gas (HRG) model fits, are summarized in Table~\ref{table2}.
\begin{table}[htbp]
\centering
\caption{Parameters for chemical freeze-out lines in Fig.~\ref{fig:4} matched to HRG model best fits \cite{Garg:2013ata}.}
\label{table2}
\begin{tabular}{ccccccc}
\toprule
     $a$ [MeV]  &  $b$ & $T_{\text{lim}}$ [MeV]  & $\mu_{\Delta}$ & $c$ & $d$ & $e$ \\ \midrule
      1785  &  0.77 & $148\sim 165$  & $-20\sim 130$ & 3.5 & 1.54  & $-0.6 $ \\ \bottomrule
\end{tabular}
\end{table}

\begin{figure*}[tbp]
\centering
\includegraphics[width=0.34\textwidth]{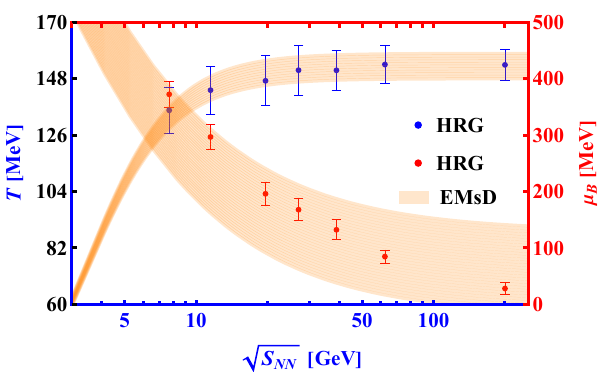}\hfill
\includegraphics[width=0.31\textwidth]{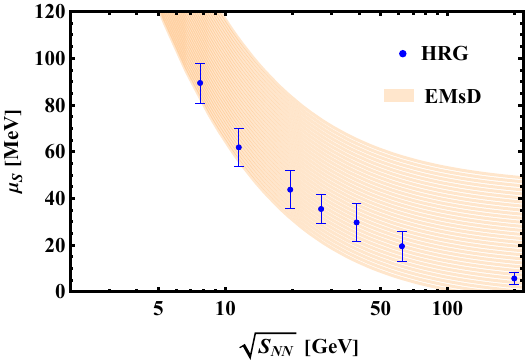}\hfill
\includegraphics[width=0.31\textwidth]{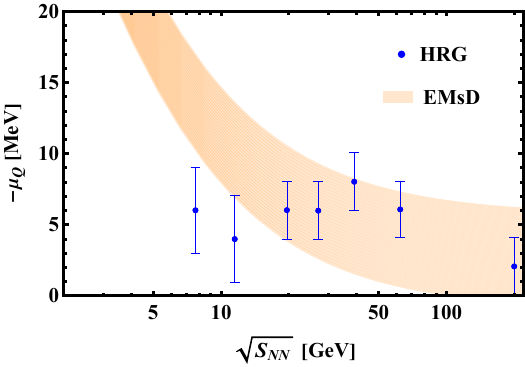}
\caption{Allowed freeze-out ranges of $T$ and $\mu_B$ (left), and self-consistently derived constrained intervals of $\mu_S$ (middle) and $\mu_Q$ (right) under experimental conditions $\rho_S=0$ and $\rho_Q/\rho_B=0.4$. The EMsD prediction bands robustly encompass the best-fit HRG parameters (data points with error bars) \cite{Garg:2013ata}. All $\mu_S$ and $\mu_Q$ bands are calculated from the $T$-$\mu_B$ freeze-out constraints in the left panel.}
\label{fig:4}
\end{figure*}
Fig.~\ref{fig:4} collects the allowed freeze-out parameter bands under net-strangeness neutrality ($\rho_S=0$) and a fixed charge-to-baryon density ratio $\rho_Q/\rho_B=0.4$. We perform the parameter fitting solely using the $T$-$\mu_B$ bands displayed in the left panel. The constrained intervals of strange chemical potential $\mu_S$ and charge chemical potential $\mu_Q$ shown in the middle and right panels are derived self-consistently from this fitted freeze-out curve. Importantly, the bounded parameter bands predicted by the EMsD model overlap well with the optimal results of the HRG model within uncertainties. This consistent matching demonstrates that our constraints imposed on all conserved charge densities based on realistic experimental freeze-out conditions are physically reasonable. Moreover, such physically motivated parameter bands guarantee that our subsequent fluctuation analysis is carried out under realistic kinematic conditions without artificial overfitting to discrete experimental data points.

\begin{figure*}[tbp]
\centering
\includegraphics[width=.31\textwidth]{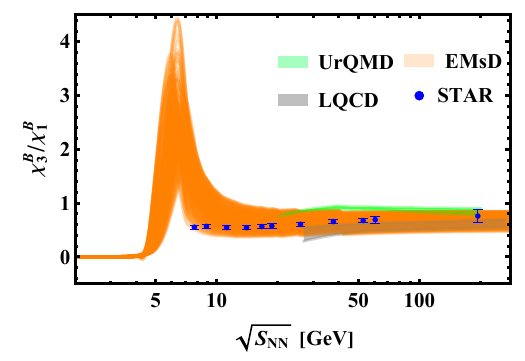}\hfill
\includegraphics[width=.31\textwidth]{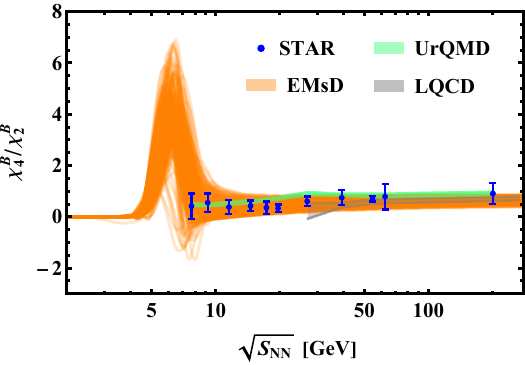}\hfill
\includegraphics[width=.31\textwidth]{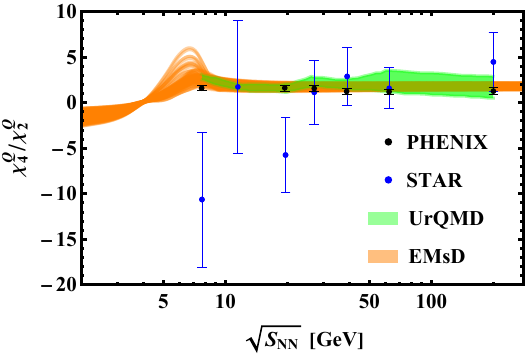}
\caption{Collision-energy dependence of fluctuation ratios: $\chi^B_3/\chi^B_1$ (left), $\chi^B_4/\chi^B_2$ (middle), and $\chi^Q_4/\chi^Q_2$ (right). Characteristic peaks originating from CEP critical fluctuations emerge within $5$--$7\,\text{GeV}$, compared with UrQMD \cite{Bass:1998ca, Bleicher:1999xi}, lattice \cite{Bazavov:2020bjn, Bollweg:2024epj}, STAR BES-I/II measurements \cite{Thader:2016gpa, STAR:2025zdq}, and PHENIX experimental data \cite{PHENIX:2015tkx}.}
\label{fig:5}
\end{figure*}
Figure~\ref{fig:5} shows the collision-energy dependence of three highly sensitive higher-order cumulant ratios for QCD critical fluctuations near the CEP: the net-baryon ratios $\chi_3^B/\chi_1^B$ and $\chi_4^B/\chi_2^B$, and the net-charge ratio $\chi_4^Q/\chi_2^Q$. Our EMsD results agree well with experimental data over the entire beam-energy range. In particular, the calculated net-baryon fluctuations faithfully reproduce the full STAR BES-II measurements, while net-charge observables match the PHENIX data, with only minor deviation from STAR BES-I at $\sqrt{s_{NN}}=7.7$ GeV.

All three fluctuation channels exhibit universal critical signatures, with well-defined peaks emerging within a narrow unified window of $\sqrt{s_{NN}}\approx 5\text{--}7$ GeV, independent of the adopted chemical freeze-out trajectories. While sharing identical peak positions, these signals feature clear hierarchical amplitudes. The net-baryon $\chi_4^B/\chi_2^B$ produces the most prominent and statistically significant critical peak, followed by a moderate signal from $\chi_3^B/\chi_1^B$, whereas the net-charge $\chi_4^Q/\chi_2^Q$ yields a considerably weaker critical structure. This hierarchy arises intrinsically from multi-charge QCD dynamics rather than numerical artifacts. Physically, net-baryon fluctuations directly couple to the CEP-associated sigma mode and develop strong critical divergence, while isospin symmetry and strangeness conservation dilute net-charge critical responses and suppress their peak amplitudes.

This hierarchical pattern provides clear guidance for ongoing experimental CEP searches. Net-baryon cumulant ratios ($\chi_4^B/\chi_2^B$, $\chi_3^B/\chi_1^B$) serve as primary high-sensitivity observables for identifying CEP signatures, and net-charge fluctuations provide an independent cross-check. The simultaneous emergence of coherent critical peaks in distinct conserved-charge channels establishes a robust dual-channel criterion to separate genuine critical fluctuations from noncritical background effects, enabling more reliable and conclusive experimental searches for the QCD critical end point.

\textit{Summary.}---
We have constructed a thermodynamically self-consistent Einstein-Maxwells-dilaton (EMsD) holographic model that intrinsically respects the grand canonical ensemble for multiple conserved charges. By enforcing exact thermodynamic relations at the holographic level, the model achieves quantitative agreement with lattice QCD at both zero and finite densities without additional tuning. We demonstrated that finite charge and strangeness densities profoundly modify the CEP landscape. Most importantly, by evaluating higher-order fluctuations along realistic freeze-out trajectories, we identified a coherent dual-channel critical peak at $\sqrt{s_{NN}} \approx 5\text{--}7\,\text{GeV}$. This provides a robust, background-free theoretical benchmark for the ongoing experimental search for the QCD critical point.

\providecommand{\href}[2]{#2}\begingroup\raggedright\endgroup

\end{document}